\documentclass[12pt]{article}
\usepackage{amsmath,amssymb}
\usepackage{epsfig}

\def\al{{\alpha^{\prime}}}

\def\bea{\begin{eqnarray}}
\def\eea{\end{eqnarray}}

\begin{document}
\thispagestyle{empty}
\vspace*{.5cm}
\noindent
DESY 05-119

\vspace*{1.6cm}

\begin{center}
{\Large\bf Gauge unification, non-local breaking, open strings}
\\[2.3cm]
{\large M. Trapletti}\\[.5cm]
{\it Deutsches Elektronen-Synchrotron, Notkestra\ss e 85,
D-22603 Hamburg, Germany} \\[.4cm]
{\small\tt (michele.trapletti@desy.de\,)}\\[1.3cm]

{\bf Abstract}\end{center}
\noindent
The issue of {\it non-local} GUT symmetry breaking is addressed
in the context of open string model building.
We study $Z_{N}\times Z_{M}^{\prime}$ orbifolds with all the
GUT-breaking orbifold elements acting freely, as rotations accompanied
by translations in the internal space.
We consider open strings quantized on these backgrounds, distinguishing
whether the translational action is {\it parallel} or {\it perpendicular} 
to the D-branes.
GUT breaking is impossible in the purely perpendicular case, 
{\it non-local} GUT breaking is instead allowed in the purely parallel case.
In the latter, the scale of breaking is set by the compactification moduli,
and there are no fixed points with reduced gauge symmetry, where
dangerous explicit GUT-breaking terms could be located.
We investigate the mixed parallel+perpendicular case in a
$Z_{2}\times Z_{2}^{\prime}$ example, having also a simplified
field theory realization. It is a new $S^{1}/Z_{2}\times Z_{2}^{\prime}$
orbifold-GUT model, with bulk gauge symmetry $SU(5)\times SU(5)$ broken
{\it locally} to the Standard Model gauge group.
In spite of the locality of the GUT symmetry breaking, there is no localized contribution
to the running of the coupling constants, and the unification scale is completely
set by the length of $S^{1}$.
\newpage

\section{Introduction}
Orbifold compactification \cite{Dixon:1985jw} provides a powerful
tool to fill the gap between string theory and particle physics.
String theory can be quantized exactly on an orbifold. The resulting model
is fully consistent, and its features are completely under control.
The orbifold action can be responsible for Supersymmetry (SUSY) and
gauge symmetry breaking. The details of the orbifold action in the
internal space/gauge bundle encode all the details of the symmetry breaking.
In particular, if an orbifold-group element (in the following ``orbifold operator'')
acts freely in the internal space, i.e. without fixed points, then the symmetry
breaking is realized {\it non-locally}, at an energy scale set by the compactification
moduli \cite{Witten:1985xc}.
Indeed, freely-acting orbifolds present a viable string embedding of 
the so-called Scherk-Schwarz symmetry breaking mechanism
\cite{Scherk:1979zr,Hosotani:1983xw}.

This mechanism has been studied in the past mainly in relation to SUSY breaking
\cite{Rohm:1983aq,Kounnas:1989dk,Kiritsis:1997ca,Antoniadis:1999ux,
Antoniadis:1999xk}. We will reconsider and use it to break gauge symmetry
\cite{Hebecker:2003we,Hebecker:2004ce}, in open string theory.
We investigate models where the gauge group of a Grand Unified Theory (GUT) is
broken, {\it non-locally}, to the Standard Model (SM) gauge group.
The scale of breaking is set by the compactification moduli, and it is 
tunable to the value predicted by the Minimal Supersymmetric Standard Model (MSSM):
$M_{GUT}=3\times 10^{16}\,{\rm GeV}$ \cite{Amaldi:1991cn}.
The compactification scale can differ from the string scale $m_{s}$, since in the resulting
model no differential running for the SM coupling constants is present between
$M_{GUT}$ and $m_{s}$.
This is the main difference between {\it non-local} breaking and
the standard {\it local} orbifold breaking. In the latter, generically,
localized contribution to the differential running can be present up to the string scale.
In this case the MSSM prediction is spoiled unless all the scales are
$\sim10^{16}\,{\rm GeV}$\footnote{The only exception is the length of the
extra dimensions unwrapped by the D-branes supporting the gauge group.}.

We describe the possible geometries that allow for non-local breaking. 
Only a small subclass of the backgrounds ensuring Scherk-Schwarz
SUSY breaking is viable, for example we select only two acceptable
backgrounds out of the $Z_{2}\times Z_{2}^{\prime}$ models classified in
\cite{Antoniadis:1999ux}.
We study $Z_{N}\times Z_{M}^{\prime}$ orbifolds, the minimal option ensuring both
non-local gauge symmetry breaking and $\mathcal{N}=1$ SUSY in 4d.
We furnish examples in the $(N,M)=(4,2)$, $(3,3)$ and $(4,4)$ cases.

These geometries guarantee non-local GUT symmetry breaking provided
that the gauge bundle has support on the whole internal space. 
In open string theory this is not ensured, since gauge bosons are localized
on D-branes that may have dimensionality lower than 10, and
extra conditions must be fulfilled.
As in the SUSY breaking case, we distinguish between {\it perpendicular} or
{\it parallel} action of the relevant (freely-acting) orbifold operators on the D-branes.
In the purely perpendicular case, the orbifold action identifies different
stacks of D-branes, and therefor cannot result in a proper GUT symmetry breaking.
In the purely parallel case, instead, such a breaking occurs.

A new interesting mechanism is realized in a mixed situation. 
The obtainable GUT symmetry breaking is local, but the differential running 
of the coupling constants due to localized contributions may be absent.
We give a $Z_{2}\times Z_{2}^{\prime}$ example with D5-branes, that
can be interpreted as a field-theory model with a single extra dimension,
compactified on $S^{1}/Z_{2}\times Z_{2}^{\prime}$.
The model has $SU(5)_{1}\times SU(5)_{2}$ bulk gauge symmetry, broken to $SU(5)$
($SU(5)^{\prime}$) in the $Z_{2}$ ($Z_{2}^{\prime}$) fixed points.
The low energy gauge symmetry is given  by the intersection of the gauge
symmetry preserved in the fixed points,  $SM=SU(5)\cap SU(5)^{\prime}$.
The breaking is then local,
but there is no differential running beyond the unification scale $M_{GUT}=R_{S^{1}}^{-1}$,
since all the localized matter must fill an $SU(5)$ ($SU(5)^{\prime}$) multiplet,
and contributes universally to the running.
This makes this model particularly interesting, since it has the simple structure
and main features of the orbifold-GUT models introduced and studied in
\cite{Kawamura:2000ev}, but it ensures an exact gauge coupling unification,
provided that the low energy spectrum is a proper one.

This mechanism could have great relevance in heterotic string model building.
As it is well known, the splitting between $M_{GUT}$ and $M_{Planck}$
can be explained by matching the unification scale with the compactification
volume.
This in general implies a string perturbativity loss \cite{Kaplunovsky:1985yy}.
In \cite{Hebecker:2004ce} it was shown that an {\it highly anisotropic} compactification
could resolve the problem, but full perturbativity would require no more than one
large extra dimension, tuned to $M_{GUT}^{-1}$, and  this is consistent (only)
with our picture.

We give here the paper outline and main results:

In Section~2 we introduce the possible background that allow for {\it non-local}
GUT symmetry breaking, furnishing two $Z_{2}\times Z_{2}^{\prime}$ examples,
a $Z_{4}\times Z_{2}^{\prime}$ example, with an equivalent $Z_{4}\times Z_{4}^{\prime}$
description, and a $Z_{3}\times Z_{3}^{\prime}$ example.

In Section~3 we study D-brane embeddings in the given backgrounds, we show
the properties of a purely perpendicular embedding (no GUT breaking), and of a purely
parallel embedding ({\it non-local} GUT breaking)\footnote{
We take a bottom-up approach and do not discuss the details of tadpole cancellation
condition, but just the details of the gauge symmetry breaking mechanisms.
To prove the existence of the described models,
at least in the $Z_{2}\times Z_{2}^{\prime}$ case,
we mainly refer to \cite{Antoniadis:1999ux}, even though we know that no ``realistic''
D-brane model of this kind have been constructed, up to now.}.
We discuss the mixed $Z_{2}\times Z_{2}^{\prime}$ case by giving a field theory
exemplification, a new five-dimensional orbifold-GUT model, with a local breaking
of the GUT symmetry but without dangerous localized contributions to the differential
running of the coupling constants.

In Section~4 we check the tree-level relations between 10d and 4d gauge and
gravity couplings.
We impose the ``observed'' values for $\alpha_{GUT}$, $M_{Planck}$
and $M_{GUT}$, we require string perturbativity and the string
scale to be larger than $M_{GUT}$, in order to avoid threshold corrections between
low energy and $M_{GUT}$.
We obtain that the described scenario is generically viable, provided that the ratio between
the string scale and $M_{GUT}$ is less than two orders of magnitude ($\lesssim 50$).

\section{GUT symmetry breaking via freely-acting \\ orbifolds}
In a freely-acting orbifold group there are operators (elements) acting
freely and operators acting non-freely. All the operators are embedded into
the gauge bundle, and their action can be GUT-preserving or GUT-breaking.
In order to have {\it non-local} GUT breaking, we require all the breaking operators
to act freely in the internal space.
We also demand that $\mathcal{N}=1$ SUSY is preserved in 4d, and that
the orbifold group is Abelian.
We clearly demand the existence of at least one GUT-breaking operator.

The minimal option fulfilling these requirements is a
$T^{6}/Z_{N}\times Z_{M}^{\prime}$ orbifold. Indeed, if we take
the generator of  $Z_{M}^{\prime}$  ($g_{M}^{\prime}$) to be freely-acting and GUT-breaking,
a non-local GUT breaking is ensured, but this is not enough.
In absence of a second orbifold action, the background geometry can always
be rewritten as a fibration of $T^{5}$ over $S^{1}$, with $\mathcal{N}\ge 2$ or
$\mathcal{N}=0$ SUSY in 4d.
An extra orbifold action is needed, to break $\mathcal{N}=2\rightarrow \mathcal{N}=1$,
and the conditions to have non-local breaking must be rechecked
on a case-by-case basis.
We take $g$, the generator of $Z_{N}$, GUT-preserving.
The requirement, than, is that the operators $g^{n} {g^{\prime}}^{m}$ ($m\neq 0$)
must act freely, since they are all GUT-breaking.

We impose that $g$ commutes with $g^{\prime}$. 
This implies that the structure of the fixed-points of $g$ must be non-trivial,
in order to have $g^{\prime}$ freely-acting. In general it is necessary the presence
of a number of $g$-fixed points multiple of the order of $g^{\prime}$.
It is easy to check that, due to this, there are no 
$Z_{2}\times Z_{3}^{\prime}$ or $Z_{3}\times Z_{2}^{\prime}$
models, and instead there are $Z_{N}\times Z_{M}^{\prime}$ examples
for  $(N,M)=(2,2),\,\,(4,2),\,\,(4,4),\,\,(3,3)$. 
Due to the fixed-point argument we conjecture the absence of models
fulfilling the requirements for ($N$, $M>6$).

\begin{figure}[t]
\hspace{1.4cm}
\includegraphics[scale=0.5]{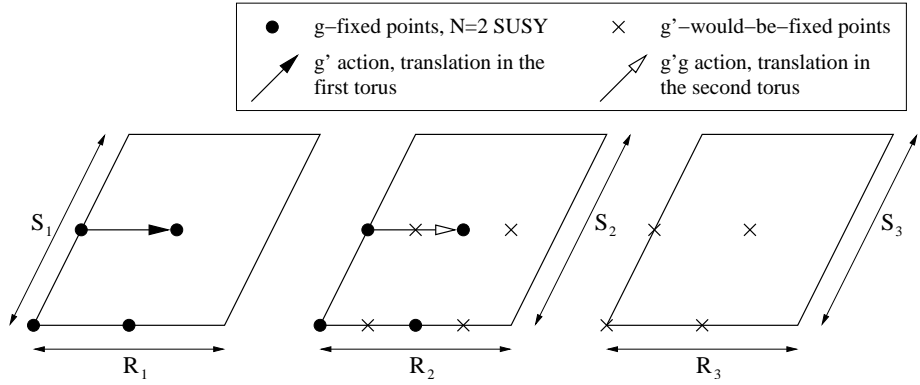}\vspace{-.3cm}
\caption{\small \it Internal geometry of the 
$Z_{2}\times Z_{2}^{\prime}$ model for $\epsilon=0$.
The action of $g$ is not free, the dots show its fixed points in the first and
second torus; since $g$ has no action in the third torus {\it each} point 
$z_{3}\in\mathbb{C}$
is a (4d) fixed point preserving $\mathcal{N}=2$ SUSY.
The arrows show the action of $g^{\prime}$ ($g\,g^{\prime}$)
in the first (second) torus: a pure translation. The crosses show the 
would-be-fixed points of $g^{\prime}$.
}\label{classz2z2}
\end{figure}

\subsection{$Z_{2}\times Z_{2}^{\prime}$ orbifold}
The freely-acting orbifolds $T^{6}/Z_{2}\times Z_{2}^{\prime}$ have been
classified in \cite{Antoniadis:1999ux}.
As introduced, we take $g^{\prime}$ GUT-breaking and $g$ GUT-preserving.
This implies that we restrict to the small subclass of models with both
$g^{\prime}$ and $g g^{\prime}$ (GUT-breaking operators) freely-acting.
For simplicity we take $T^{6}=T^{2}_{1}\times T^{2}_{2}\times T^{2}_{3}$.
We parametrize each $T^{2}_{i}$ with $z_{i}\in\mathbb{C}$ having periodicity
$z_{i}\sim z_{i}+(R_{i} n+\tau_{i} S_{i} m)$, $|\tau|=1$.
Imposing the requirement we reduce
the classification of  \cite{Antoniadis:1999ux} to the two models
\cite{Hebecker:2003we,Hebecker:2004ce}
\bea
g:\left\{
	\begin{array}{l}
	z_{1}\rightarrow -z_{1}\\
	z_{2}\rightarrow -z_{2}\\
	z_{3}\rightarrow  z_{3}+\epsilon R_{3}/2
	\end{array}\right.&&
g^{\prime}:\left\{
	\begin{array}{l}
	z_{1}\rightarrow z_{1}+R_{1}/2\\
	z_{2}\rightarrow -z_{2}+R_{2}/2\\
	z_{3}\rightarrow -z_{3}
	\end{array}\right.
\eea
with $\epsilon=0,\,1$. In both cases $g^{\prime}$ and $g g^{\prime}$ are freely-acting,
since their action is a translation along the real part of $z_{1}$ and $z_{2}$ respectively.
If $\epsilon=0$, $g$ is not freely-acting, and there are fixed points (planes) of
reduced ($\mathcal N=2$) SUSY, see Fig.~\ref{classz2z2}. If instead
$\epsilon=1$, also $g$ is freely-acting, and each point in the internal space
is $\mathcal N=4$ supersymmetric. The latter configuration is very similar
to the smooth Calabi Yau case.

Taking $g$ GUT-preserving and $g^{\prime}$ GUT-breaking and imposing that the gauge bundle
has support on the plane parametrized by the real part of $z_{1}$ and $z_{2}$,
we obtain that the breaking scale is $M_{GUT}^{2}=R_{1}^{-2}+R_{2}^{-2}$.
It is then necessary to tune $R_{1}^{-1},\,R_{2}^{-1}\sim 10^{16}\,{\rm GeV}$, while the
other radii are free.
Due to the absence of $\mathcal{N}=1$ fixed points, it could be difficult
to introduce chiral matter in complete representations of the unified gauge group.
The problem can be solved by introducing stacks of intersecting D-branes
\cite{Berkooz:1996km}, as in the case discussed by \cite{Watari:2004jg}, and putting
the unified group on one of the stacks.

\begin{figure}[t] 
\hspace{1.5cm}
\includegraphics[scale=0.5]{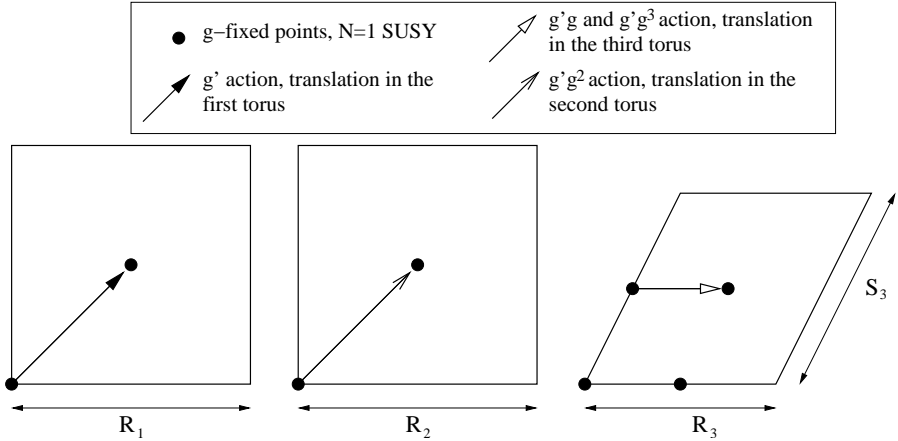}\vspace{-.3cm}
\caption{\small \it
Internal geometry of the $Z_{4}\times Z_{2}^{\prime}$ model.
The action of $g$ is not free, the dots show the 4d fixed points.
They preserve $\mathcal{N}=1$ SUSY.
The arrows show the action of $g^{n}\,g^{\prime}$, a pure translation
in the first (n=0), second (n=2) or third torus (n=1,\,3).
}\label{classz4z2}
\end{figure}

\subsection{$Z_{4}\times Z_{2}^{\prime}$  ($Z_{4}\times Z_{4}^{\prime}$) orbifold}
We found only one $Z_{4}\times Z_{2}^{\prime}$ (abelian) model compatible with the requirements.
The toroidal geometry is constrained, due to the crystallographic action of $Z_{4}$:
$\tau_{i}=i$, $R_{i}=S_{i}$, for $i=1,\,2$.
The orbifold action is
\bea
g:\left\{
	\begin{array}{l}
	z_{1}\rightarrow e^{i\pi/2} z_{1}\\
	z_{2}\rightarrow e^{i\pi/2} z_{2}\\
	z_{3}\rightarrow -z_{3}
	\end{array}\right.&&
g^{\prime}:\left\{
	\begin{array}{l}
	z_{1}\rightarrow z_{1}+R_{1}(\tau_{1}+1)/2\\
	z_{2}\rightarrow -z_{2}+R_{2}(\tau_{2}+1)/2\\
	z_{3}\rightarrow -z_{3}+R_{3}/2
	\end{array}\right.
\eea
All the combinations $g^{n} g^{\prime}$ are freely-acting, as required:
$g^{\prime}$ is a diagonal translation in the first torus, $g^{2n+1} g^{\prime}$ is
a diagonal translation in the third torus, $g^{2} g^{\prime}$ is
a diagonal translation in the second torus (see Fig.~\ref{classz4z2}).
The Dp-brane embedding issue is discussed in detail in the next section.
We anticipate that a complete non-local GUT breaking is ensured only when
the gauge bundle fills all the directions where the orbifold action is a translation.
The scale of breaking is, in that case, related to the volume of these
directions.
In within this picture, we have to set $R_{i}^{-1}\sim 10^{16}\, {\rm GeV}$
for all $i$'s, and the only free volume parameter is $S_{3}$.

The described $Z_{4}\times Z_{2}^{\prime}$ model can also be seen 
as a  $Z_{4}\times Z_{4}^{\prime}$ model. Indeed,
taking an action
\bea
g:\left\{
	\begin{array}{l}
	z_{1}\rightarrow e^{i\pi/2} z_{1}\\
	z_{2}\rightarrow e^{i\pi/2} z_{2}\\
	z_{3}\rightarrow -z_{3}
	\end{array}\right.&&
g^{\prime}:\left\{
	\begin{array}{l}
	z_{1}\rightarrow e^{i\pi/2} z_{1}+R_{1}(\tau_{1}+1)/2\\
	z_{2}\rightarrow e^{-i\pi/2}z_{2}+R_{2}(\tau_{2}+1)/2\\
	z_{3}\rightarrow z_{3}+R_{3}/2
	\end{array}\right.
\eea
it is possible to check that the generated orbifold group is the same as in
the $Z_{4}\times Z_{2}^{\prime}$
orbifold, but each operator can be obtained in two different ways. In particular notice that
$g^{2} {g^{\prime}}^{2}\equiv I$ in the internal space.
This imply that $g^{2} {g^{\prime}}^{2}$ must be embedded as the identity operator
also in the gauge bundle, and that the action of ${g^{\prime}}^{2}$
must be GUT-preserving, since $g$ is GUT-preserving.

\begin{figure}[t]
\hspace{0cm}
\includegraphics[scale=0.46]{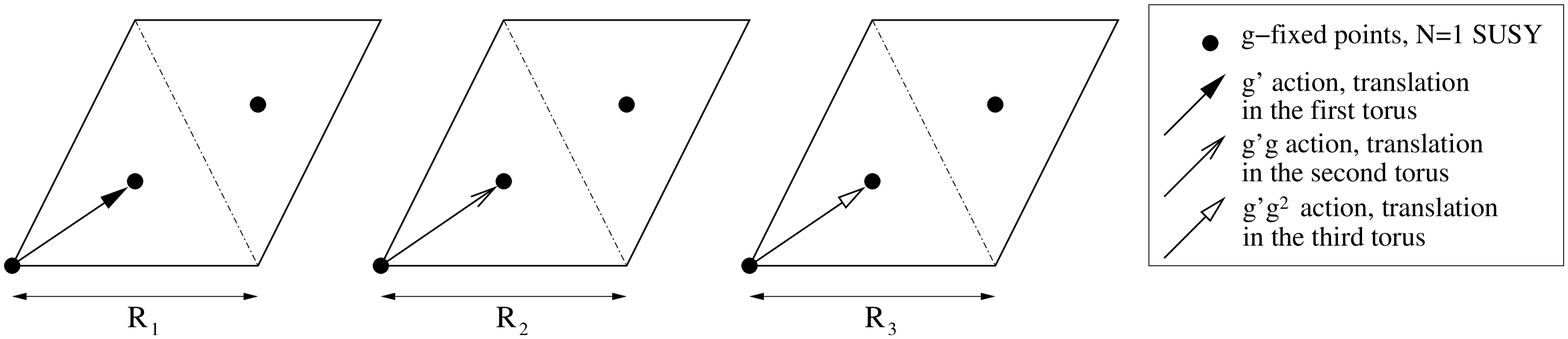}\vspace{-.3cm}
\caption{\small \it
Internal geometry of the $Z_{3}\times Z_{3}^{\prime}$ model.
The action of $g$ is not free, the dots show the 4d fixed points.
They preserve $\mathcal{N}=1$ SUSY.
The arrows show the action of $g^{n}\,{g^{\prime}}^{m}$, a pure translation
in the first (n=0), second (n=m) or third torus (n=2m).
}\label{classz3z3}
\end{figure}

\subsection{$Z_{3}\times Z_{3}^{\prime}$}
As in the previous case the complex structure is (completely) fixed
by the orbifold action, $\tau_{i}=e^{i\,\pi/3}$,
$R_{i}=S_{i}$, for all $i$'s.
The orbifold action is
\bea
g:\left\{
	\begin{array}{l}
	z_{1}\rightarrow e^{-4\pi i/3} z_{1}\\
	z_{2}\rightarrow e^{2\pi i/3} z_{2}\\
	z_{3}\rightarrow e^{2\pi i/3} z_{3}
	\end{array}\right.&&
g^{\prime}:\left\{
	\begin{array}{l}
	z_{1}\rightarrow z_{1}+R_{1}(\tau_{1}+1)/3\\
	z_{2}\rightarrow e^{-2\pi i/3} z_{2}+R_{2}(\tau_{2}+1)/3\\
	z_{3}\rightarrow e^{2\pi i/3} z_{3}+R_{3}(\tau_{3}+1)/3
	\end{array}\right.
\eea
Notice that the orbifold group is abelian, as required.
The action of $g^{n}\,{g^{\prime}}^{m}$ is always free for $m\neq 0$, as required:
for $n=0$ it is diagonal translation in the first torus, for $n=m$ a diagonal
translation in the second,
for $n=2m$ a diagonal translation in the third (see Fig.~\ref{classz3z3}).

\begin{figure}[t]
\hspace{0.7cm}
\includegraphics[scale=0.5]{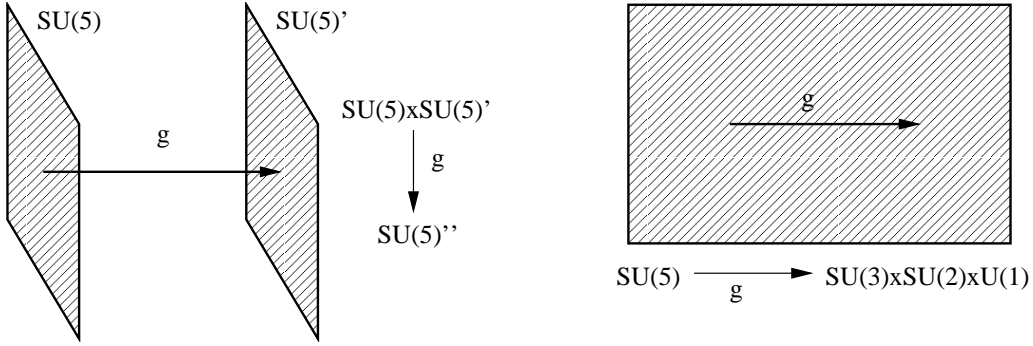}\vspace{-.3cm}
\caption{\small \it
Gauge symmetry breaking due to a freely-acting operator $g$
acting as a translation in a direction {\it perpendicular} (on the left) or
{\it parallel} to the D-brane (on the right).
In the first case no ``proper'' GUT breaking is realized, but a rank
reduction of the gauge group. In the second case the possibility of
a non-local GUT symmetry breaking is ensured.
}\label{orthogonal}
\end{figure}

\section{Open string model building}
The described geometries ensure a {\it non-local} GUT
symmetry breaking, but it is necessary that the orbifold elements are
embedded into the gauge bundle as specified.
The requirement is that the GUT symmetry breaking operators
must be present and act freely.

In heterotic string theory these requirements can be always fulfilled,
provided that modular invariance conditions are also fulfilled.
In open string theory, instead, it could be impossible to embed  a freely-acting
operator in a GUT-breaking way.

In open string theory the gauge bosons live on Dp-branes, not necessarily filling the
whole spacetime ($p$ can be less than $9$). A freely-acting operator can act as a translation
in a direction that is {\it parallel} or {\it perpendicular} to the relevant stack
of Dp-branes (See Fig.~\ref{orthogonal}).

In the perpendicular case, the point where the Dp-branes reside is mapped
into a different point, where a new stack of Dp-branes must be located.
Consistency requires the two Dp-brane stacks to be completely equivalent,
with the same gauge group $\mathcal G$. The orbifold action identifies the two groups,
and only a (diagonal) combination is left invariant: $\mathcal G\times \mathcal 
G\rightarrow \mathcal G$.  A ``proper'' GUT symmetry breaking is impossible,
but a rank reduction
occurs (for details check, in the $Z_{2} \times Z_{2}$ case, 
\cite{Antoniadis:1999ux,Clements:2004wp}).
This implies that if {\it all} the freely-acting operators act in a perpendicular way no
GUT breaking at all can be embedded.

In the parallel case, the Dp-brane stack (gauge group) is mapped onto
itself, and a GUT breaking can be realized. If all the freely-acting operators
act in a parallel way, then no rank reduction is realized. We can take the generator
of $Z_{N}$ to be GUT-preserving, the generator of $Z_{M}^{\prime}$ to be GUT-breaking,
and we have a non-local GUT symmetry breaking due to the orbifold elements
$g^{n}\,{g^{\prime}}^{m},\,m\neq 0$. The scale of breaking is  given by the volume of the
dimensions where $g^{n}\,{g^{\prime}}^{m}$ act as a translation.
Clearly, it is necessary also to fulfill the tadpole cancellation conditions, that impose
strong constraints on the kind of gauge groups and spectra that can be present.
We do not introduce the issue here, we only notice that it was undertaken,
in the $Z_{2}\times Z_{2}^{\prime}$ case, in \cite{Antoniadis:1999ux}.
In particular a non-local GUT symmetry breaking of the described kind can be realized,
the unifying group being $U(16)$ broken to $U(16-n)\times U(n)$.
The gauge group is not realistic, but the situation can be highly improved by introducing
continuos Wilson lines. We leave for future work a detailed study of tadpole 
cancellation condition and realistic model building.

These gauge symmetry breaking patterns are closely related to the kind of SUSY breaking
studied, for example, in \cite{Antoniadis:1999ux,Angelantonj:2000xf},
but the fate of SUSY in a freely-acting orbifold is generically different than the 
fate of gauge symmetry.
As an example, since bulk SUSY is always present, 
a freely-acting orbifold is responsible in any case for a {\it parallel} SUSY breaking,
at least in the closed string sector.
This does not happen in the gauge symmetry case.

\begin{figure}[t]
\hspace{0.7cm}
\includegraphics[scale=0.5]{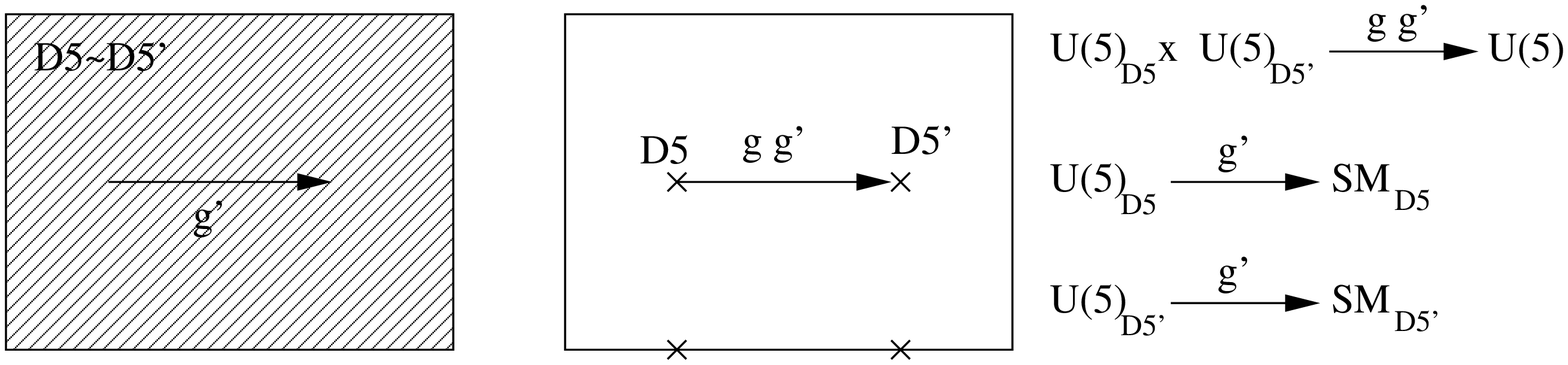}\vspace{-.3cm}
\caption{\small \it
The background described in Fig.~\ref{classz2z2} with two stacks of 5 D5-branes
filling the first torus and located in two would-be-fixed points of $g^{\prime}$.
The gauge group before of the orbifold action is generically 
$U(5)_{D5}\times U(5)_{D5^{\prime}}$,
the action of $g^{\prime}$ is parallel to the two stacks, and can be embedded in a 
GUT-breaking way. The action of $g g^{\prime}$ is instead perpendicular and induces an
identification
of the two $U(5)$. 
}\label{z2z2seminonlocal}
\end{figure}

Between the two extreme cases (complete perpendicular/parallel case) we have
the intriguing possibility of an intermediate case. Since the orbifold group
contains more than one freely-acting operator, we may have a stack of Dp-branes parallel
to the action of an  operator and perpendicular to the action of another one.
We illustrate this possibility with a $Z_{2}\times Z_{2}^{\prime}$ example. 
We consider the geometry of Fig.~\ref{classz2z2} and embed two stacks of 5 D5-branes, filling
the first torus and localized into two different would-be-fixed points of $g^{\prime}$,
in the second torus, as shown in Fig~\ref{z2z2seminonlocal}.
The gauge group, in absence of any orbifold projection, is $U(5)_{D5}\times U(5)_{D5^{\prime}}$.
The action of $g^{\prime}$ is parallel to both the two stacks, and can be embedded in a 
GUT-breaking way. The action of $g g^{\prime}$ is instead perpendicular, and induces an 
identification
between the two $U(5)$'s. The ending unbroken group is then just a diagonal $U(3)\times U(2)$
combination.
The breaking is {\it local}, as we show in the following, but the unification properties are
particularly interesting, and deserve some extra comment.
We explain them by introducing a simplified field theory example.

\begin{figure}[t]
\hspace{3.7cm}
\includegraphics[scale=0.6]{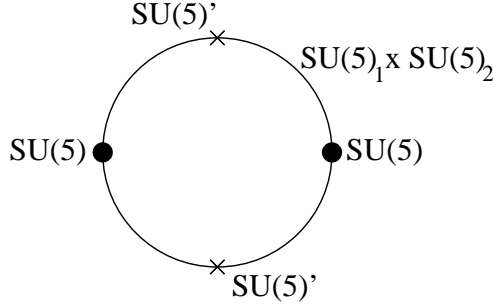}\vspace{-.3cm}
\caption{\small \it
An $S^{1}/Z_{2}\times Z_{2}^{\prime}$ field theory model.
The bulk symmetry is $SU(5)_{1}\times SU(5)_{2}$, broken to
$SU(5)$ and to $SU(5)^{\prime}$ in the $g$ and $g^{\prime}$
fixed points (dots and crosses respectively).
The surviving gauge group is just the SM gauge group:
$SU(5)\cap SU(5)^{\prime}=SU(3)\times SU(2)\times U(1)$.
 }\label{esseuno}
\end{figure}

\subsection{A field theory model on $S^{1}/Z_{2}\times Z_{2}^{\prime}$}
From a purely field theoretical point of view, the described D5-brane construction
is completely reproduced by a model with a single extra dimension,
parametrized by $x\in [0,2\pi R)$.
We take the extra dimension to be compactified on $S^{1}/Z_{2}\times Z_{2}^{\prime}$.
The action of the $Z_{2}$ operator is $g:\,x\rightarrow -x$, with fixed points $0$
and $\pi R$, while the action of the $Z_{2}^{\prime}$ operator is
$g^{\prime}:\,x\rightarrow -x+\pi R$, with fixed points $\pi R/2$ and $3\pi R/2$
(see Fig.~\ref{esseuno}).
We introduce a bulk gauge group $SU(5)_{1}\times SU(5)_{2}$, we embed $g$
in the gauge group such that $g:SU(5)_{1}\leftrightarrow SU(5)_{2}$, more precisely,
defining $T^{ab}_{i}$ a generator of $SU(5)_{i}$\footnote{We take $a,\,b=1,2\dots,5$, and
define $T^{aa}$ as the 5 Cartan generators of $U(5)$, from which we exclude the ``diagonal''
generator $\sum_{a} T^{aa}$.} the identification is
$g:T^{ab}_{1}\leftrightarrow T^{ab}_{2}$, and the surviving gauge group in $0$ and $\pi R$
is a {\it diagonal} $SU(5)$ generated by $T^{ab}_{1}+ T^{ab}_{2}$\footnote{Obviously
it is crucial the presence, from the very beginning, of a $Z_{2}$ symmetry linking
the two $SU(5)_{i}$, that must have, for example, the same coupling constant.
This is ensured by consistency of the freely-acting orbifold in the string theory case,
since we require the two stacks of D5-branes to be symmetric}.
We embed then $Z_{2}^{\prime}$ action as 
$g^{\prime}:T^{ab}_{1}\rightarrow \delta^{a}_{c}T^{cd}_{2}\delta^{b}_{d}$
with 
\bea
\delta=\left(
		\begin{array}{ccccc}
		1&0&0&0&0\\
		0&1&0&0&0\\
		0&0&1&0&0\\
		0&0&0&-1&0\\
		0&0&0&0&-1
		\end{array}\right).
\eea
In this way the surviving gauge group in $\pi R/2$ and $3\pi R/2$ is a {\it different}
$SU(5)^{\prime}$, and the intersection of the two gauge groups is just the
SM gauge group. The breaking is local, since the $SU(5)$ symmetry preserved in $0$
is generically broken in $\pi R/2$, but the SM differential running of the coupling constants
is generated only by the bulk degrees of freedom, and it stops precisely
at the unification scale $M_{GUT}=R^{-1}$.
There is no fixed points contribution to the differential running since only
full multiplets of $SU(5)$ ($SU(5)^{\prime}$) can be localized there,
and the SM is embedded exactly in the {\it same} way in $SU(5)$ and in $SU(5)^{\prime}$.
The last point is crucial: a multiplet of $SU(5)$ behaves as a multiplet of $SU(5)^{\prime}$ 
under the action of the SM gauge group, and they both contribute universally to
the running of the SM coupling constants.
We expect the same argument to be valid for any localized contribution to the action.

We have shown, in the $Z_{2}\times Z_{2}^{\prime}$ model, that also a D-brane embedding
that does not ensure non-local breaking can ensure the required phenomenology,
i.e. no localized contribution to the differential running of the coupling constants.
We expect this issue to be valid even for the other geometries, even though we do
not furnish here a case-by-case proof.

This mechanism could have great relevance in heterotic string model building,
where complete string perturbativity is ensured {\it only} for
a single extra dimension tuned to
$M_{GUT}=3\times 10^{16}\,{\rm GeV}$, and the others taken at the inverse string scale
\cite{Hebecker:2004ce}, that is fixed close to the Planck mass.
An embedding of the described effect would be a model where string
perturbativity is not spoiled and
$M_{GUT}$  is introduced as an intermediate (compactification) scale
between the the string scale
and the low energy scales \cite{Kobayashi:2004ud}, without any need for large threshold
corrections at the string scale.
Nevertheless, in the mechanism a rank reduction is needed for the gauge group, and a standard
abelian orbifold configuration is not enough. It would be necessary to introduce non-abelian
freely-acting orbifolds/continuos Wilson lines \cite{Ibanez:1987xa} (for recent studies
see also \cite{Forste:2005rs}).

\section{Energy scales}
The gravitational interactions are described, in Type II string theory,
by the following 10d bosonic action (see \cite{Hebecker:2004ce} 
for notation and conventions)
\bea
S_{Grav}=-\int d^{10}x \sqrt{G} \frac{1}{2\,(2\pi)^{7}\al^{4}g^{2}}R,
\eea
where $G$ is the 10d metric, $R$ the Ricci scalar, $\al$ is
the string scale and $g$ is the string coupling.
Using these definition for the 10d coupling constants, the
gauge interactions, due to the presence of a Dp-brane stack, are described
by the action
\bea
S_{Dp}=\int dx^{p+1}\sqrt{*G} \frac{1}{8\,(2\pi)^{p-2}\,
\al^{\frac{p-3}{2}}g}{\rm Tr_{V}}F^{2},
\eea
where $*G$ is the metric induced on the Dp-brane stack.
The string scale is related to the mass of the first excited state as $m_{I}=\sqrt{\al}$,
$g$ has been chosen in such a way that the weak coupling regime is defined for $g\ll1$
(boundary at $2$).
This can be proven either by a duality argument (S-duality with heterotic
theory for $p=9$, as an example), or simply computing
the ratio between a generic one-loop amplitude correction to a given tree-level 
quantity\footnote{The computation is done introducing $m_{I}$ as UV cutoff
of the $p+1$ dimensional field theory,  and considering an $SO(32)$ gauge group.
For further details see \cite{Hebecker:2004ce}.}
\bea
\frac{\rm one\,\,loop}{\rm tree\,\,level}=\frac{2 (2\pi)^{p-2}}
{(p-3) 2^{p+1} \pi^{\frac{p+1}{2}}\Gamma\left(\frac{p+1}{2}\right)}
30 g=N(p)\,g,
\eea
and checking that $N(p)$ is always $O(1)$.

The relations between these couplings and the 4d couplings
depend on the total compactification volume $V$ 
(gravitational coupling) and on the volume of the space filled
by the Dp-brane $V_{//}$ (YM coupling).
We take a trivial compactification on a space given by the 
direct product of six circles. The circles have radius $R_{//}$,
if the Dp-brane wraps them, or $R_{\bot}$ if they are not wrapped.
Consequently we have
$V_{//}=(2\pi R_{//})^{p-3}$ and $V=V_{//}\times (2\pi R_{\bot})^{9-p}$,
and the 4d Planck mass and gauge coupling are, respectively
\bea
\label{coup1}
M_{P}^{2}=\frac{4 R_{\bot}^{9-p} R_{//}^{p-3}}{\al^{4} g^{2}},\,\,\,\,\,\,
\alpha_{GUT}=\frac{\al^{(p-3)/2}}{R_{//}^{p-3}}g.
\eea 
We define dimensionless radial parameters $r=R/\sqrt{\al}$ and
rescaled dimensionless volumes $v_{//}=r_{//}^{p-3}$, $v_{\bot}=r_{\bot}^{9-p}$.
We also introduce $m_{I}=1/\sqrt{\al}$, the mass of the first-excited string state.
Then, Eq.~(\ref{coup1}) can be rewritten, rescaling $g\rightarrow 2\,g$, as
\bea
\label{treerel}
M_{P}^{2}=\frac{v_{\bot}v_{//}}{g^{2}}m_{I}^{2},\,\,\,\,\,\,
\frac{\alpha_{{GUT}}}{2}=\frac{g}{v_{//}}.
\eea
The second equation states the usual impossibility
of a large Dp-brane volume, i.e. the impossibility
of $R_{//}\gg1/m_{I}$. Nevertheless, notice that  
$m_{I}$ is not fixed close to the Planck mass as in
the heterotic case.

\subsection{Viable configurations}
The tree-level relations of Eq.~(\ref{treerel}) state a distinction
between the volume filled by the D-brane and the volume
orthogonal to the D-brane.
We should introduce a further distinction, and specify the directions
where the orbifold elements act as a translation.
We can have then four different classes of radii/volumes.
Between them, we consider the class of the directions wrapped by the D-branes AND
where the orbifold action is free. These dimensions set the unification scale and
must have inverse radius $\sim 3\times 10^{16}\,{\rm GeV}$.
All the other dimensions are generically unfixed.
Taking the number of the dimensions with radius $M_{GUT}^{-1}$ to be $d$
we can rewrite  Eq.~(\ref{treerel}) as
\bea
\label{lasteq}
\left(\frac{M_{P}}{m_{I}}\right)^{2}
\left(\frac{M_{GUT}}{m_{I}}\right)^{d}
g^{2}=r_{\bot}^{9-p} r_{//}^{p-3-d},\,\,\,\,
\left(\frac{M_{GUT}}{m_{I}}\right)^{d} g=
\frac{\alpha_{GUT}}{2} r_{//}^{p-3-d},
\eea
where we take, with abuse of notation, $r_{//}$ to be the radii of the dimensions wrapped
by the D-branes and where the orbifold action is not free.

Now we can impose constraints on the values of the various parameters,
and check the consistent possibilities.
Our guideline is a model with just the MSSM running of
the coupling constants between the low energy scale
($M_{Z}$) and the unification scale $M_{GUT}=3\times10^{16}\,
{\rm GeV}$, where the unified gauge group is restored and
the couplings meet. This implies, in Eq.~(\ref{lasteq})
(i) $m_{I}>M_{GUT}$;
(ii) $r_{//}<m_{I}/M_{GUT}$, to avoid light charged Kaluza-Klein modes;
(iii) $r_{\bot}>M_{GUT}/m_{I}$, to avoid light charged winding modes.
We also impose $r_{//}>1$, to avoid the presence of light winding modes
for the gravitational part of the action, that could be annoying in a low
energy field theory (SUGRA) description of the model: in presence of ``short'' radii
we consider always a T-dual description.

The constraints are very mild. We distinguish between the $p=9$ and $p<9$ cases.
In the first case there are no orthogonal tunable radii. More precisely, the  
equations can be cast as (substituting the values for $\alpha_{GUT}
$ etc.)
\bea
r_{//}^{6-d}\sim\frac{1}{66} \left(\frac{M_{GUT}}{m_{I}}\right)^{d-2},\,\,\,\,\,
g=50\left(\frac{m_{I}}{M_{P}}\right)^{2}.
\eea
The condition $M_{GUT}<m_{I}$ implies $r_{//}<1/3$ for the minimal case
$d=2$, and even smaller values for $d>2$. Since we prefer to avoid the presence
of ``short'' radii we always consider T-dual versions with $p<9$.

For $p<9$ we can always rewrite
Eq.~(\ref{lasteq}) as
\bea
 \left(\frac{M_{GUT}}{m_{I}}\right)^{-1}=(50 g)^{1/d} r_{//}^
 {-\frac{p-3-d}{d}},\,\,\,\,r_{\bot}^{9-p}=
 \frac{g}{50}\left(\frac{M_{P}}{m_{I}}\right)^{2}.
\eea
The requirement $M_{GUT}<m_{I}$, $r_{//}>1$
and $g<1$ imply $\frac{1}{50}<g<1$.
This means that, if $r_{//}\sim 1$, the ratio between $M_{GUT}/m_{i}$
is bounded between $1$ ($g=1/50$) and $(1/50)^{1/d}$ ($g=1$), i.e.
that in the relevant cases $d=1,\,2,\,3$ the maximal hierarchy
between $m_{I}$ and $M_{GUT}$ is, respectively (roughly), $50,\,7,\,4$.
An higher hierarchy can be obtained only relaxing the requirement $r_{//}>1$.
The constraints on $r_{\bot}$ are also always fulfilled, with $r_{\bot}$ typically large.

\section*{\large Acknowledgments}\vspace{-.2cm}
I am grateful to Wilfried Buchm\"uller, Stefan Groot Nibbelink,
Marco Serone, Alexander Westphal and especially Arthur Hebecker
for discussions and comments.

\end{document}